\documentclass{vgtc}                          % final (conference style)
\graphicspath{{figures/}{pictures/}{images/}{./}} % where to search for the images

\usepackage{times}                     % we use Times as the main font
\usepackage{tabu}                      % only used for the table example
\usepackage{booktabs}                  % only used for the table example
\usepackage{lipsum}                    % used to generate placeholder text
\usepackage{mwe}                       % used to generate placeholder figures

\usepackage{mathptmx}                  % use matching math font
\usepackage[svgnames]{xcolor}
\usepackage[frozencache,cachedir=.]{minted}

\definecolor{darkbackground}{rgb}{0.1,0.1,0.1}
\definecolor{darkframe}{rgb}{0.2,0.2,0.2}

\setminted{
  fontsize=\small,
  breaklines=true,
  breakanywhere=true,
  tabsize=2,
  linenos=false,
  frame=single,
  framesep=1mm,       % Reduced frame separation
  xleftmargin=0pt,    % No left margin within the code block
  xrightmargin=0pt,   % No right margin within the code block
  style=default,
  bgcolor=white
}

\onlineid{1056}

\vgtccategory{Research}

\vgtcinsertpkg

\title{Semantic Scaffolding: Augmenting Textual Structures with Domain-Specific Groupings for Accessible Data Exploration}

\author{%
\begin{tabular}{@{}c@{\hspace{2em}}c@{\hspace{2em}}c@{}}
Jonathan Zong\thanks{Co-first authors} \,\thanks{e-mail: jzong@mit.edu} & 
Isabella Pedraza Pineros\footnotemark[1] \,\thanks{e-mail: ipedraza@mit.edu} & 
Mengzhu (Katie) Chen\thanks{e-mail: mzc219@mit.edu} \\
{\scriptsize MIT CSAIL} & 
{\scriptsize MIT CSAIL} & 
{\scriptsize \parbox{1.4in}{\centering MIT CSAIL}} \\
\end{tabular}\\[1ex]
\begin{tabular}{@{}c@{\hspace{4em}}c@{}}
Daniel Hajas\thanks{e-mail: d.hajas@ucl.ac.uk} & 
Arvind Satyanarayan\thanks{e-mail: arvindsatya@mit.edu} \\
{\scriptsize Global Disability Innovation Hub} & 
{\scriptsize \parbox{1.4in}{\centering MIT CSAIL}} \\
\end{tabular}
}

\teaser{
  \centering
  \includegraphics[width=\linewidth]{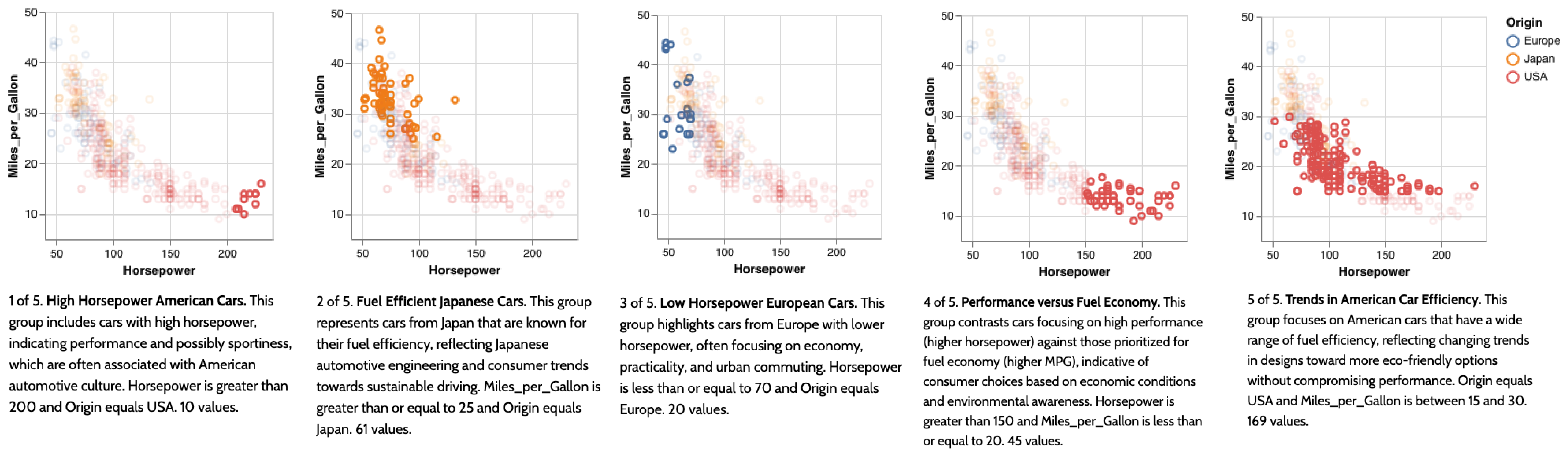}
  \caption{Data highlights generated from an example cars dataset using the \textit{semantic scaffolding} technique. Each highlight is shown with a visualization of its query predicate, demonstrating that highlights correspond to areas of interest in the data.}
  \label{fig:teaser}
}

\abstract{
    Drawing connections between interesting groupings of data and their real-world meaning is an important, yet difficult, part of encountering a new dataset.
A lay reader might see an interesting visual pattern in a chart but lack the domain expertise to explain its meaning.
Or, a reader might be familiar with a real-world concept but struggle to express it in terms of a dataset's fields.
In response, we developed \textit{semantic scaffolding}, a technique for using domain-specific information from large language models (LLMs) to identify, explain, and formalize semantically meaningful data groupings.
We present groupings in two ways: as \textit{semantic bins}, which segment a field into domain-specific intervals and categories; and \textit{data highlights}, which annotate subsets of data records with their real-world meaning.
We demonstrate and evaluate this technique in Olli, an accessible visualization tool that exemplifies tensions around explicitly defining groupings while respecting the agency of readers to conduct independent data exploration.
We conducted a study with 15 blind and low-vision (BLV) users and found that readers used semantic scaffolds to quickly understand the meaning of the data, but were often also critically aware of its influence on their interpretation.

} % end of abstract


\begin{document}


\firstsection{Introduction}

\maketitle

When a lay reader encounters a new dataset on a website or news article, an important part of their exploratory process is to identify interesting groupings of data, and explain them in terms of that data's real-world meaning.
Broadly, the real-world meaning of data is known as its \textit{semantics}; domain-specific knowledge about data is how a reader determines whether a number ``represent[s] a day of the month, or an age, or a measurement of height, or a unique code for a specific person, or a postal code for a neighborhood, or a position in space'' \cite{munzner_visualization_2014} or other possible meanings.
Effectively drawing connections between data and its semantics can be challenging for a lay reader.
For instance, a reader might see an interesting visual pattern in a chart but lack the domain-specific knowledge to explain its meaning.
In this case, difficulty arises because the reader does not know what they don't know --- it may be difficult to know how to begin to acquire the relevant context.
Or, a reader might be familiar with a real-world concept but struggle to translate it into the terms of the dataset.
For example, a reader may not know how to approximate a subjective concept like ``sports car'' in terms of fields like \texttt{Horsepower} and \texttt{Miles\_per\_Gallon} --- or possibly other fields in the dataset that were not in the initial visualization.
These challenges can affect a reader's comprehension of the data --- as prior research has shown, a reader's interpretation of a visualization is sensitive to differences in their prior knowledge and personal background \cite{hullman_visualization_2011, peck_data_2019}.

In response, we developed a technique called \textit{semantic scaffolding} for using domain-specific information from large language models (LLMs) to identify, explain, and formalize semantically meaningful data groupings.
Rather than use an LLM to generate summary descriptions or chat responses, we use it to return groupings as data structures that include a \textit{name}, \textit{explanation}, and \textit{query predicate}.
For instance, in the example cars dataset, an LLM might create a grouping named ``Fuel Efficient Japanese Cars'' with the following explanation: ``This group represents cars from Japan that are known for their fuel efficiency, reflecting Japanese automotive engineering and consumer trends towards sustainable driving'' (\autoref{fig:teaser}).
Crucially, the LLM associates this grouping with the following query predicate: \texttt{Miles\_per\_Gallon} $\geq$ \texttt{25} $\cap$ \texttt{Origin} $=$ \texttt{Japan}.
This allows a reader to connect the name and explanation with an explicitly-defined subset of data.
We developed two types of interface elements for presenting semantic groupings to a reader: \textit{semantic bins}, which segment a single field (i.e. column) into
domain-specific intervals and categories; and \textit{data highlights}, which annotate subsets of data records (i.e. rows) with their real world meaning.
These two uses of semantic scaffolding in an interface serve distinct purposes.
Semantic bins help a reader break down a single field into understandable pieces, facilitating navigation and exploration through a dataset.
Data highlights help a reader quickly get an overview of a dataset, indicating potentially interesting subsets to begin exploring further.

We prototyped these designs via extensions to Olli \cite{blanco_olli_2022}, an accessible visualization system for screen reader users.
Data accessibility is a rich context in which to evaluate our work because current tools for blind and low vision (BLV) screen reader users amplify the challenges lay readers of visualizations face.
First, sighted readers often rely on their visual perception to identify interesting groupings in a visualization, but screen reader interfaces rarely afford a similar type of overview.
Second, BLV readers highly value the agency to independently explore data, conduct an open-ended interpretive process, and arrive at their own conclusions about the data \cite{lundgard_accessible_2022}.
For instance, research has shown that screen reader users find descriptions less useful when they over-emphasize contextual and domain-specific information at the expense of descriptive statistics or the data values themselves \cite{lundgard_accessible_2022, choe_enhancing_2024}.
These two factors require screen reader interfaces to strike a balance between making groupings available without overly prescribing a reader's interpretation.
In our preliminary user testing, we found that textual scaffolding was able to accelerate BLV readers' understanding of a dataset's real-world meaning, but that readers were often critically aware of AI's influence on their interpretation process.

\section{Related Work}

\subsection{Accessible Textual Data Representations}

To make data visualizations accessible to screen reader users, a designer must provide descriptions that can be read as text-to-speech.
Because conventional static alt text does not afford data exploration at varying levels of detail comparable to strategies sighted readers employ, researchers have turned to \textit{structured textual descriptions} \cite{zong_rich_2022}, which enable screen reader users to navigate along a hierarchy and move between overview and detail with textual descriptions.
Accessible visualization systems that incorporate structured textual descriptions include Olli \cite{blanco_olli_2022}, Data Navigator \cite{elavsky_data_2023}, VizAbility \cite{gorniak_vizability_2023}, Chart Reader \cite{thompson_chart_2023}, and Umwelt \cite{zong_umwelt_2024}.
This work introduces a technique for augmenting textual structures with LLM-generated groupings, and demonstrates the technique via extensions to Olli.

\subsection{Generating Textual Descriptions with Language Models}

Natural language generation of textual descriptions for data visualization is an area of research that has received renewed attention due to recent advances in large language models (LLMs).
There are generally two interface design approaches that these systems have taken.
First are systems that provide a chat-like interface with which a user can query a description by inputting a question in natural language (often known as chart question answering systems) \cite{kavaz_chatbot-based_2023, shen_towards_2023, gorniak_vizability_2023}.
Second are systems that generate standalone summary descriptions of charts, similar in format to conventional alt text written by humans.
This work includes systems like Chart-to-Text \cite{obeid_chart--text_2020} and DataTales \cite{sultanum_datatales_2023}.
The advantage of using LLMs to generate descriptions is that they can automatically incorporate contextual and domain-specific information --- also known as L4 semantic content \cite{lundgard_accessible_2022} --- into descriptions \cite{kim_exploring_2023}.
However, a critical limitation of this approach is the possibility that generated descriptions can contain errors, including hallucinations \cite{ji_survey_2023}.
But even if generated captions were correct all of the time, there would still be drawbacks to existing approaches.
For instance, Choe et al. noted that users sometimes become over-reliant on the LLM in a chart question answering system instead of developing their own interpretations of the data \cite{choe_enhancing_2024}.
Similarly, summarization-based approaches have the same limitations that conventional alt texts do; namely, that they lack affordances for information granularity and limit users' ability to conduct self-guided exploration \cite{zong_rich_2022}.
In our work, we introduce a new technique --- distinct from Q\&A or summarization --- for using LLMs to incorporate domain-specific information.
We use an LLM to generate semantically meaningful groupings, and use that output to support structural navigation \cite{zong_rich_2022}.
\section{Semantic Scaffolding: Identifying, Explaining, and Formalizing Meaningful Data Groupings}

Semantic scaffolding is a technique for using domain-specific information from a large language model (LLM) to guide a reader's understanding of a dataset's meaning.
We engaged in an iterative co-design process involving a blind co-author in which we prototyped methods for incorporating domain-specific information into a user interface for data exploration.
This prototyping process revealed different types of user needs that semantic scaffolding could address, which we explore via two types of interface elements: \textit{semantic bins} and \textit{data highlights}.

\subsection{Semantic Bins}
Binning is a common operation for analyzing and communicating data that involves dividing a field into equally-sized intervals that cover the extent of the field's data values.
Most data visualizations implicitly use binning to generate axis ticks, and accessible textual data representations frequently use binning to structure a screen reader's navigation through data.

The conventional approach to binning does not take into account domain-specific information; computing equally-sized bins is a function that can be applied to any dataset.
However, lack of semantic information could make it more difficult for a reader to understand and contextualize a field's data values.
For example, in a dataset about cars, a bin function might segment the \texttt{miles\_per\_gallon} field by equally-sized increments of 10.
However, a reader might not know what range of values is considered a low vs. high fuel efficiency, or whether an increment of 10mpg represents a large or small difference in fuel efficiency. 
As a result, it might be difficult for them to map the numbers onto their subjective understanding of fuel efficiency.

\begin{figure}[htb]
  \centering
  \includegraphics[width=\linewidth]{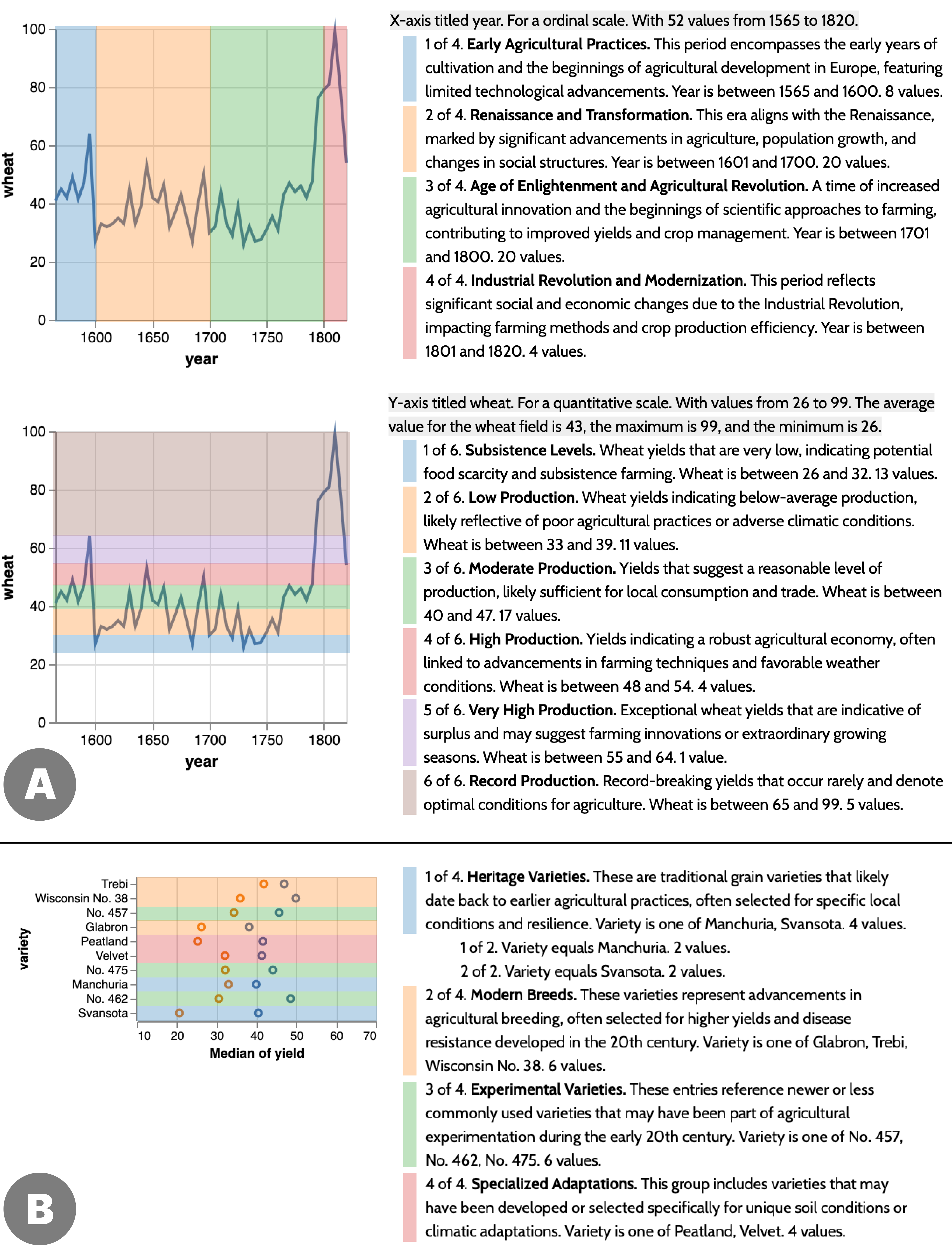}
  \caption{\label{fig:semantic-bin}
           Semantic binning using a example wheat and barley datasets. A) the \texttt{year} field (temporal) is binned into historical periods, and \texttt{wheat} yield (quantitative) is binned into levels of low to high production. B) barley \texttt{variety} (nominal) is grouped into higher-level variety types.}
\end{figure}

Semantic bins are groupings that use domain-specific information to segment a field into higher-level intervals and categories that express a dataset's meaning.
\autoref{fig:semantic-bin} shows an example usage of semantic binning applied to fields with a variety of measure types (temporal, quantitative, and nominal).
For a continuous (e.g., temporal, quantitative) field, we prompt an LLM to re-bin the field, specifying in the prompt that bins should be non-overlapping intervals that cover the extent of the data.
In the figure, the semantic bins map onto how a reader might make sense of years (via historical periods corresponding to agricultural practices), and amount of wheat yield (via levels from low to high relative to the typical or expected yield) (\autoref{fig:semantic-bin}A).
For a categorical field, we prompt an LLM to group categories into higher-level groupings, specifying in the prompt that the groupings should be mutually exclusive and exhaustively cover all categories.
The figure example takes the names of barley varieties and groups them into meaningful high-level categories, such as ``Heritage Varieties'' or ``Modern Breeds'' (\autoref{fig:semantic-bin}B).

\subsection{Data Highlights}
Data highlights are groupings of data records along criteria that correspond to a real-world interpretation.
In contrast to a semantic bin, which is defined by a predicate involving only one field, a data highlight's predicate can involve multiple fields to select a subset of data records.
For example, each data highlight in \autoref{fig:teaser} represents a semantically-meaningful subset of cars in the example car dataset, with a predicate that involves two or more of the fields \texttt{Horsepower}, \texttt{Miles\_per\_Gallon}, and \texttt{Origin}.

Data highlights are akin to visual annotations for readers, which are a technique that designers frequently use to emphasize and draw attention certain parts of the data \cite{ren_chartaccent_2017}.
Indeed, data highlights have the same components as many instances of visual annotation: a defined subset of data records, and a semantically-meaningful explanation.
However, we think of data highlights as independent of any specific visual representation.
In our examples, we convey data highlights both as a conditional encoding in a visualization (e.g. visually annotating the included data points using color or opacity), and as descriptions in a textual structure (\autoref{fig:teaser}).

Data highlights are designed to help lay readers understand a dataset even if they lack prior knowledge about the data domain.
For example, in the cars dataset from \autoref{fig:teaser}, a reader might observe the inverse relationship between \texttt{Horsepower} and \texttt{Miles\_per\_Gallon} but lack the context to know why there might be a tradeoff between the two, or how cultural factors between the three \texttt{Origin} values contribute to differences.
In \autoref{fig:teaser}, each data highlight focuses on a different region of the chart, and connects the selected data points with a real-world explanation.
For example, the third data highlight in the figure is called ``Low Horsepower European Cars,'' and explains the tendency for European cars to have lower horsepower in terms of economy, practicality, and urban commuting.
The highlight includes a query predicate defining the grouping's inclusion criteria in terms of the \texttt{Horsepower} and \texttt{Origin} fields, allowing the reader to connect the explanation to concrete values in the dataset.

\subsection{Limitations}
\label{sec:limitations}

\begin{figure}[htb]
  \centering
  \includegraphics[width=\linewidth]{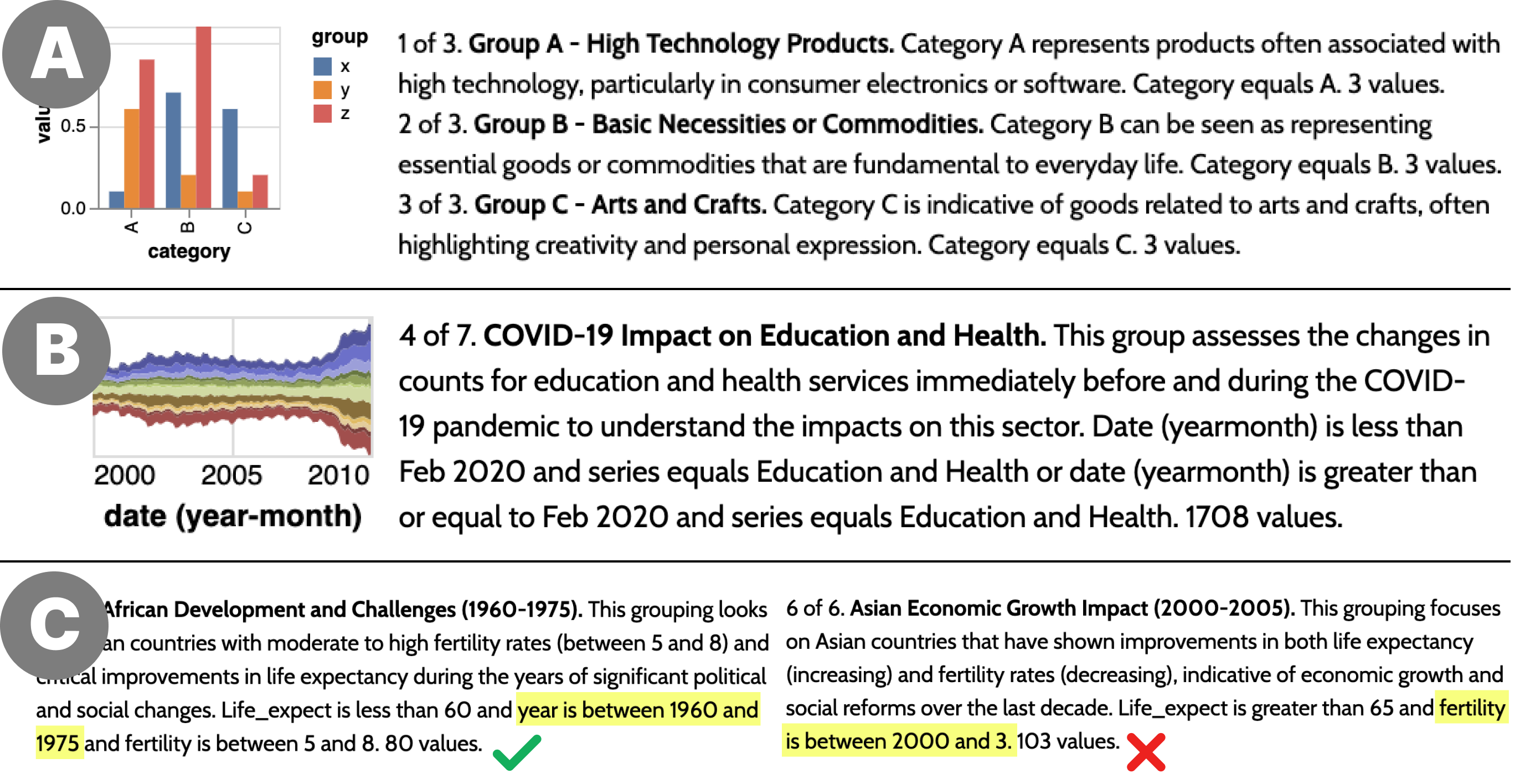}
  \caption{\label{fig:errors}
           Examples of LLM errors encountered in our design process. A) LLM hallucinates dataset semantics when there is not enough information. B) LLM includes information that is outside of the dataset's context. C) LLM generates incorrect query predicate.}
\end{figure}

Because semantic scaffolding is a technique that relies on LLMs, it is limited by the kinds of errors LLMs tend to make.
Here, we provide examples of errors we observed in our prototyping process.

\textit{Hallucinated dataset semantics.}
When a dataset does not have meaningful field labels, the LLM might hallucinate a domain for the data.
In \autoref{fig:errors}A, the data has the generic field labels of \texttt{category} and \texttt{value}.
Even though there's no information about what these fields mean, the LLM interprets them as consumer products.

\textit{Social context inappropriate to dataset.}
The LLM may incorporate domain-specific information that is not necessarily factually wrong, but is outside of the scope of the dataset.
\autoref{fig:errors}B shows a data highlight referencing the COVID-19 pandemic, which started in 2020, for an unemployment dataset that only covers 2000--2010.

\textit{Incorrect query predicates.}
\autoref{fig:errors}C demonstrates an error where the LLM identified a reasonable grouping, but was unable to formulate a correct query predicate that matches the explanation. The figure shows one correct predicate and one incorrect predicate generated from the same data. 
While the \texttt{fertility} between \texttt{2000} and \texttt{3} term is syntactically valid, it is semantically incorrect.

\textbf{Discussion of limitations.}
In our prototyping process, we explored recent techniques for LLM output validation, such as SymGen \cite{hennigen_towards_2024}.
However, we found that most data-to-text validation techniques were not appropriate for our use case, because we are not merely asking the LLM to summarize a dataset.
In our case, our goal is to leverage the LLM's knowledge that is external to the dataset.
As a result, validating this information requires either access to an external knowledge base, or use of a proxy like the model's internal confidence \cite{tian_fine-tuning_2024}. 
Therefore, because this is still an active area of research, we were not able to incorporate reliable technical approaches to validating the output of semantic scaffolding at this time. However, in our user testing, we surfaced insights on user behaviors and strategies for validating and thinking critically about LLM output  (\autoref{sec:eval}).

\section{Implementation}

We implemented semantic scaffolding by prompting \texttt{gpt-4o-mini} with instructions to create and return data groupings, attaching a full dataset to the prompt.
To ensure that we receive a structured response from the LLM, we defined a data structure that specifies our expected response format.
Each grouping identified by the LLM is returned as a data structure with three properties: a name that summarizes the content of the grouping; a longer explanation of the grouping's meaning; and a query predicate that defines the criteria for a data point to be included in the grouping.
For compatibility with existing visualization tools, the query predicate is specified in Vega-Lite's existing predicate syntax \cite{satyanarayan_vega-lite_2017}.
\autoref{fig:type-def} contains a definition of the data structure in TypeScript syntax.

{\vspace{-\baselineskip}\noindent
\begin{figure}[H]     %
\setlength{\abovecaptionskip}{0pt}
\setlength{\belowcaptionskip}{0pt}

\begin{minted}{typescript}
type LLMResponse = {
  groups: SemanticScaffold[];
};

type SemanticScaffold = {
  name: string;
  explanation: string;
  predicate: LogicalComposition<FieldPredicate>;
};

{
  "name": "AAPL Price Surge During the Tech Boom",
  "explanation": "This group focuses on Apple's stock price during the late 2000s and early 2010s, highlighting its rise in value matched with the smartphone revolution and innovation of products like the iPhone.",
  "predicate": {
    "and": [
      { "field": "symbol", "equal": "AAPL" },
      { "field": "price", "gte": 150 },
      { "field": "date", "range": 
        [ "2008-08-31", "2012-12-31" ] }
    ]
  }
}
\end{minted}
\caption{\label{fig:type-def}Type definitions and example for semantic scaffolding}
\end{figure}
}

\section{Evaluation}
\label{sec:eval}

We conducted an exploratory study with 15 blind and low-vision (BLV) participants who use screen readers. The goal of this evaluation was to form an initial understanding of how our technique affects readers' data exploration and interpretation.
The study involved 100-minute Zoom interviews with 15 BLV participants exploring prototype implementations of semantic scaffolding in Olli \cite{blanco_olli_2022}.
Here, we report some qualitative themes that emerged from our initial observations.

\textbf{Semantic scaffolding helps readers understand and contextualize data.}
When encountering a new dataset, readers need to understand the data's real-world meaning first.
P1 illustrated this point, saying ``the numbers don't mean much until I know a little bit about what I'm reading about.''
We found that participants were able to quickly identify what the data refers to using semantic scaffolds even if their initial understanding was incorrect.
For instance, participants had many different initial guesses about the prototypes' data semantics.
Based on the \texttt{Flipper Length (mm)} and \texttt{Body Mass (g)} fields, some participants initially thought the penguins dataset referred to fish (P1), seals (P2), or dolphins (P3).
However, all these participants correctly updated their understanding to reflect information about penguins found in the data highlights.
Accelerating this process of identifying data semantics can be very important to a reader's experience: as one participant put it, ``I don't feel dumb looking at this data'' (P4).

One participant contrasted semantic scaffolding positively with prior AI-based tools they had used.
P9, who had used chart question-answering tools in the past, noted that ``sometimes you don’t know what questions to ask'' when you are not familiar with the data.
Further, they said ``sometimes its unfair to expect you to ask the right questions because you don’t have a clue of [...] what the data is about.''
Semantic scaffolding can address this ``cold start'' problem when participants start with limited prior knowledge.

\textbf{Readers critically appraise semantic scaffolds using their prior knowledge.}
Semantic scaffolds prompted participants to think about data in semantically meaningful ways, incorporating their own prior knowledge.
For example, P8 noted that ``car manufacturers don't [categorize cars 
with] engine sizes, they just base it on fuel economy,'' and suggested ``utility'' as a potential additional category.
In this case, the participant had additional context or expertise that made the LLM's categorization insufficient for them, even if it was not technically wrong.

Sometimes, participants were skeptical of arbitrariness and asked for justification.
P11 questioned the subjectivity of certain categories: ''I don't know how arbitrary the cutoff is. Like, what would you say a gas guzzler is? That's maybe arbitrary, maybe not. I don't know.''
Even when participants lacked domain knowledge to fully evaluate explanations, they were still aware of potential limitations in the groupings.

Willingness to verify the output was context-dependent.
Many said that they would more likely check against other sources if they were looking at data for professional purposes (P1), making a presentation (P2), or if it was otherwise important (P6, P8).
In our study, participants used the data table to compare against the description of the grouping's query predicate (P11).

\textbf{BLV readers also want sighted readers to have semantic scaffolding, to preserve common ground.}
For many participants, the use of semantic scaffolding made them reflect on situations where they are collaborating with sighted people.
For example, participants noted that semantic scaffolding provides ``maybe more than a sighted person would get from such a graphic because it gets you info about the relevance of the values with knowledge you don't have'' (P6) and that this ``feels weird when working with sighted people'' (P8).
Participants generally felt that it was important to maintain parity of information access with sighted people.
For example, P10 asked, ``does the visual person get the same info?''
Prior work has emphasized the importance of common ground between mixed-ability collaborators \cite{zong_umwelt_2024}, and participants' concerns reflected the idea that it was important to them to participate in and lead conversations about data even in spaces that have a majority of sighted people.
P14 explained, ``whenever you can put
sighted and blind individuals on the same playing field with equal footing, you've accomplished something that's
really marvelous. [...] For us to be able to be literally on the
same page, it's helpful because we're looking at the very same information.''
For P14, accessibility is a two-way street: ``[blind people] have to
accommodate [visual people] in a way.''
It was important to them to include sighted collaborators in their data analysis process, even when using assistive tools like semantic scaffolding.
These discussions led us to reflect on the distinction in our work between \textit{disability inclusion for BLV people} and \textit{making data understandable for lay readers}, two ideas of accessibility that can be complementary and mutually-reinforcing.

\section{Conclusion}

In this paper, we introduced semantic scaffolding: a technique for using an LLM to generate meaningful groupings of data to aid lay readers in understanding unfamiliar datasets.
With semantic scaffolding, readers benefit from domain-specific knowledge that is incorporated into descriptions and navigational structures.
We instantiated this approach in Olli, and observed that users find semantic scaffolds valuable for forming an initial understanding of a dataset while applying their prior knowledge to evaluating the trustworthiness of descriptions.
Our work suggests a use for LLMs to augment user interfaces with domain-specific affordances for interpretation.

\acknowledgments{
This work was supported by NSF \#2341748, MIT-SERC, and MIT-Google.}

\bibliographystyle{abbrv-doi}

\bibliography{access-llm}
\end{document}